\documentclass{aastex}          
\usepackage{spr-astr-addons}    

\RequirePackage{color}

\begin{document}
%
\title{Pulsation Period Changes as a Tool to Identify Pre-Zero Age Horizontal Branch Stars}
\shorttitle{Pre-Zero Age HB Stars}
\shortauthors{Silva Aguirre et al.}

\author{Silva Aguirre, V.} 
\affil{Max Planck Institute for Astrophysics, Garching, Germany}
\email{vsilva@mpa-garching.mpg.de} 
\and 
\author{Catelan, M.}
\affil{Departamento de Astronom\'ia y Astrof\'isica, Pontificia Universidad Cat\'olica de Chile}
\and 
\author{Weiss, A.}
\affil{Max Planck Institute for Astrophysics, Garching, Germany}
\and 
\author{Valcarce, A.A.R}
\affil{Departamento de Astronom\'ia y Astrof\'isica, Pontificia Universidad Cat\'olica de Chile}


\begin{abstract}
One of the most dramatic events in the life of a low-mass star is the He flash, which takes place at the tip of the red giant branch (RGB) and is followed by a series of secondary flashes before the star settles into the zero-age horizontal branch (ZAHB). Yet, no stars have been positively identified in this key evolutionary phase, mainly for two reasons: first, this pre-ZAHB phase is very short compared to other major evolutionary phases in the life of a star; and second, these pre-ZAHB stars are expected to overlap the loci occupied by asymptotic giant branch (AGB), HB and RGB stars observed in the color-magnitude diagram (CMD).
We investigate the possibility of detecting these stars through stellar pulsations, since some of them are expected to rapidly cross the Cepheid/RR Lyrae instability strip in their route from the RGB tip to the ZAHB, thus becoming pulsating stars along the way. As a consequence of their very high evolutionary speed, some of these stars may present anomalously large period change rates. We constructed an extensive grid of stellar models and produced pre-ZAHB Monte Carlo simulations appropriate for the case of the Galactic globular cluster M3 (NGC 5272), where a number of RR Lyrae stars with high period change rates are found. Our results suggest that some -- but certainly not all -- of the RR Lyrae stars in M3 with large period change rates are in fact pre-ZAHB pulsators.
\end{abstract}


%

\section{Introduction}\label{s:1}

Low-mass stars, after exhausting their central hydrogen supply, develop increasingly more massive helium cores due to the continued supply of He-rich material from the H-burning shell that surrounds it. Eventually, the conditions in the core become ripe for the ignition of the He-burning triple-alpha reactions. Since the core at this point is electron-degenerate, the process does not take place quiescently, but rather in the form of a thermonuclear runaway, called He flash. In this process, the star will typically change its luminosity by several orders of magnitude; the surface temperature may also increase very significantly. The inner structure of the star undergoes dramatic changes in the process as well. And yet, in spite of the importance of this evolutionary phase, to the best of our knowledge not a single such star has positively been identified to date.
The main goal of the present study is, accordingly, to perform the first systematic study of the expected properties of pre-ZAHB stars, with a view towards their detection using a combination of pulsation and color-magnitude diagram (CMD) properties. For more details, see \citet{ref:vs08}.

\section{Stellar Grid and Population Synthesis Code}\label{s:2}

\begin{center}
\begin{figure}[h]
\includegraphics[width=\columnwidth]{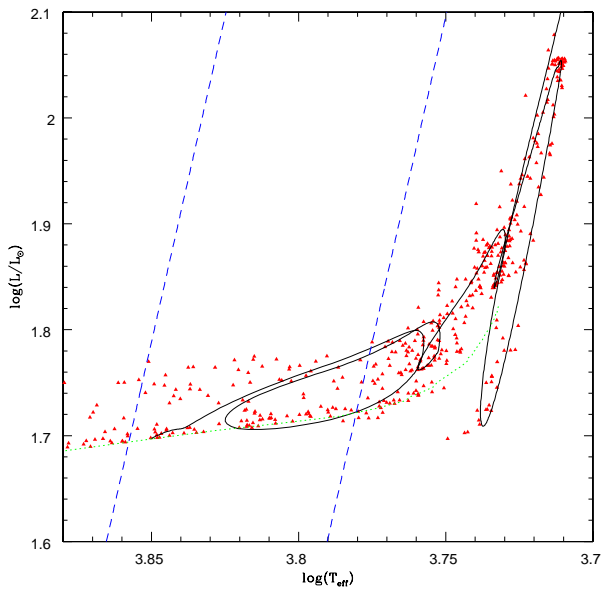}
\caption{Results for one simulation with 457 pre-ZAHB stars. Red triangles indicate our simulated stars, the solid line depicts the evolutionary track for $\eta_{\rm{R}}$ = 0.41, and the green dotted line shows the position of the ZAHB for different mass values. Blue dashed lines are the instability strip boundaries (schematic).} 
\label{fig:1}
\end{figure}
\end{center}

We used the Garching Stellar Evolution Code \citep{ref:ws08} to construct the evolutionary tracks required to populate the pre-ZAHB phase. We adopted the mixing length theory with a value $\alpha_{\rm{MLT}}$ = 1.75, and a chemical composition of Z = 0.001 and Y = 0.25, as appropriate for M3. Our models were computed with an initial mass on the zero-age main sequence (ZAMS) of 0.85 M$_{\odot}$ and application of mass loss in its most widely used formulation, namely the Reimers (\citeyear{ref:dr75,ref:dr77}) formula,

\begin{center}
\begin{equation}\label{eqn:1}
\frac{\partial \rm{M}}{\partial \rm{t}}=-4 \times 10^{-13} \eta_{\rm{R}} \frac{L}{gR}\,\,\, M{_\odot} \rm{yr}^{-1},
\end{equation}
\end{center}

\noindent where $L$ is the stellar luminosity (in solar units), $g$ the surface gravity (in cgs units), and $R$ the surface radius. A thin interpolation grid was created by evolving stars from the ZAMS to the ZAHB for a range of $\eta_{\rm{R}}$ from 0.0 to 0.60.
We adopted a Gaussian-shaped mass distribution at the tip of the RGB \citep{ref:vc08}), with mean value $<\rm{M}>$ = 0.647 M $_{\odot}$ and $\sigma$ = 0.022 M$_{\odot}$. We generated the masses for the stars populating the pre-ZAHB randomly, and interpolated over the tracks to obtain their luminosity and temperature. We then computed the blue edge of the instability strip (IS) as in \citet{ref:fc87}, after applying a correction of -200 K. The width of the IS has been taken as $\Delta\rm{log(T_{eff})}$ = 0.075. Periods and period change rates are calculated for the stars contained within the IS using the formulation of \citet{ref:fc98}. In Fig. \ref{fig:1} we present the result of one of our simulations with one of our evolutionary tracks overplotted.

\section{Empirical Data}\label{s:3}
Magnitudes and colors for the variable stars in M3 were obtained from \citet{ref:cc05}. For the non-variable stars, we used the photometry provided by \citet{ref:rb94} and \citet{ref:ff97}. M3 has $\sim$530 HB stars whose position in the CMD have been measured. If we consider lifetimes, pre-ZAHB evolution lasts $\sim$1.4 Myr, while the HB phase takes $\sim$80 Myr. Thus, one should expect to find one pre-ZAHB star for every 60 or so bona fide HB stars.

\section{Results}\label{s:4}
\begin{figure}[h]
 \includegraphics[width=\columnwidth]{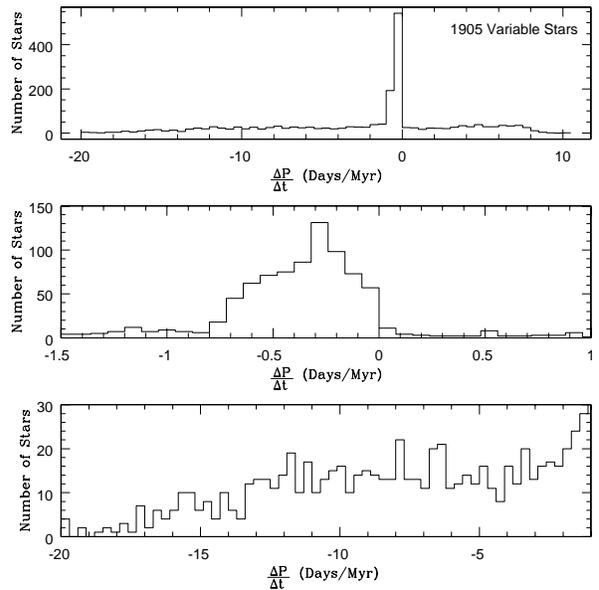}
 \caption{Distribution of the period change rates. The middle panel zooms in around the peak of the distribution, while the lower panel shows the wing of the distribution for the more extreme (negative) values.} 
\label{fig:2}
 \end{figure}

In Fig. \ref{fig:2} we present a histogram that shows the expected period change rate distribution for an M3-like input mass distribution. As can be seen, most of the pre-ZAHB stars have period change rates contained in the [-1,0] d/Myr range, meaning that evolution occurs mainly towards higher temperatures and/or decreasing luminosity, as expected for these types of stars.

%

Therefore, among 530 HB stars in M3, approximately 9 should be in the pre-ZAHB phase -- two of which being pulsating (RR Lyrae-like) stars. We emphasize the fact that $\Delta P/\Delta t$ absolute values higher than approximately 0.1 d/Myr are completely unexpected on the basis of canonical evolution, but have often been observed among field and cluster RR Lyrae stars.

\begin{figure}[h]
 \includegraphics[width=\columnwidth]{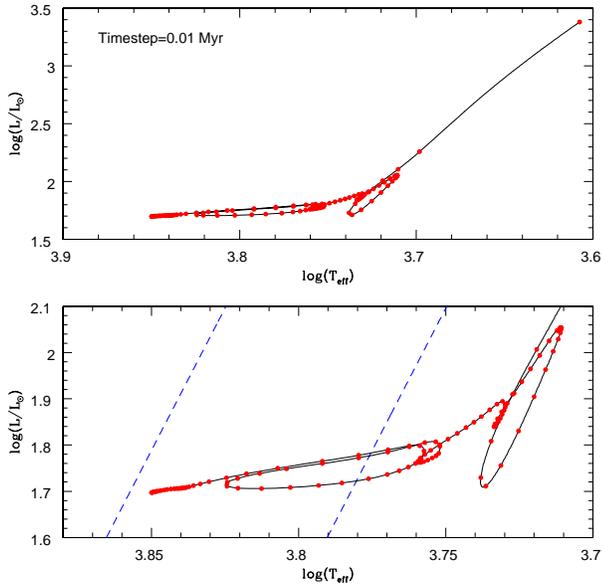}
 \caption{Pre-ZAHB evolution on the Hertzsprung-Russell diagram for the $\eta_{\rm{R}}$ = 0.41 evolutionary track. Red dots are plotted for every 0.01 Myr of evolution since the main flash.} 
 \label{fig:3}
 \end{figure}

In Fig. \ref{fig:3}, the time distribution along the evolutionary track corresponding closely to the most likely mass among M3's HB stars is presented. A red dot was placed at every 0.01 Myr of evolution, starting from the He flash down to the ZAHB.

%

As can be seen, the most extreme negative values of the period change rate can be obtained during the (very) fast evolution along the secondary loops. However, this is a relatively unlikely situation, since pre-ZAHB stars spend most of their time in the blueward evolution that takes place after the last secondary flash (just prior to the ZAHB proper). This implies that most of the pre-ZAHB variables will have negative period change rates in the aforementioned [-1,0] d/Myr range.

\begin{figure}[h]
 \includegraphics[width=\columnwidth]{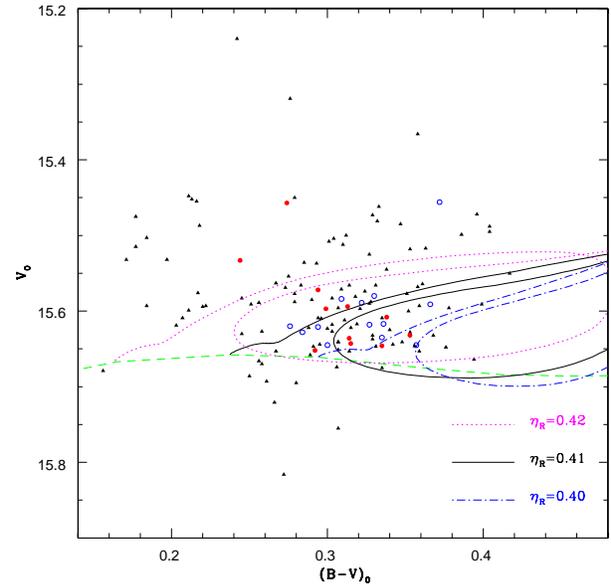}
 \caption{Distribution of the M3 RR Lyrae stars on the CMD. Filled red circles: stars with $\Delta\rm{P}/\Delta\rm{t} < -0.15$ d/Myr; open blue circles: stars with $\Delta\rm{P}/\Delta\rm{t} > +0.15$ d/Myr; triangles: RR Lyrae stars with smaller (or undetermined) $\Delta\rm{P}/\Delta\rm{t}$ values.} 
 \label{fig:4}
 \end{figure}

In Fig. \ref{fig:4}, we present the RR Lyrae stars of M3 with three of our evolutionary tracks overplotted, whose associated mass values at the tip of the RGB make them representative of the majority of the pre-ZAHB stars in the cluster. As can be seen, the positions of these variables in the CMD, as well as their period change rate values, are broadly consistent with at least some of them being pre-ZAHB stars. Indeed, according to our simulations, at least $\sim$2 of them very likely are pre-ZAHB pulsators.

%

\section{Conclusions}\label{s:5}
Our pre-ZAHB Monte Carlo simulations show that, for an M3-like HB mass distribution, the pre-ZAHB CMD positions should be basically indistinguishable from those of bona fide HB stars in the cluster. We predict that, in the specific case of M3, $\sim$9 pre-ZAHB stars should be present, $\sim$2 of which most likely are disguised as RR Lyrae stars with relatively large (and negative) period change rates.

\acknowledgments Support for M.C. is provided by Proyecto Basal PFB-06/2007, by FONDAP Centro de Astrof\'{i}sica 15010003, and by Proyecto FONDECYT Regular \#1071002.
%

\end{document}